# Determination of Personalized Asthma Triggers from Evidence based on Multimodal Sensing and Mobile Application

Revathy Venkataramanan[1], Dipesh Kadariya[1], Hong Yung Yip[1], Utkarshani Jamini[1], Krishnaprasad Thirunarayan[1], Maninder Kalra[2], Amit Sheth[1]

[1] Ohio Center of Excellence in Knowledge-enabled Computing (Kno.e.sis),Department of Engineering and Computer Science, Wright State University, Dayton, OH, USA
[2] Dayton Children's Hospital, Dayton, OH, USA

Corresponding Author: Amit Sheth, PhD, Ohio Center of Excellence in Knowledge-enabled Computing (Kno.e.sis),377 Joshi Research Center, Wright State University, 3640 Col. Glenn Hwy,Dayton, OH 45435, USA (amit@knoesis.org)

Word Count: 3508

## ABSTRACT
### Objective
Asthma is a chronic pulmonary disease with multiple triggers manifesting as symptoms with various intensities. This paper evaluates the suitability of long-term monitoring of pediatric asthma using diverse data to qualify and quantify triggers that contribute to the asthma symptoms and control to enable personalized management plan.

### Materials and Methods
Asthma condition, environment, and adherence to the prescribed care plan were continuously tracked for 97 pediatric patients using kHealth-Asthma technology for one or three months.

### Result
At the cohort level, among 21% of the patients deployed in spring, 63% and 19% indicated pollen and Particulate Matter (PM2.5), respectively, as the major asthma contributors. Of the 18% of the patients deployed in fall, 29% and 21% found pollen and PM2.5 respectively, to be the contributors. For the 28% of the patients deployed in winter, PM2.5 was identified as the major contributor for 80% of them. One patient across each season has been chosen to explain the determination of personalized triggers by observing correlations between triggers and asthma symptoms gathered from anecdotal evidence.

### Discussion and Conclusion
Both public and personal health signals including compliance to prescribed care plan have been captured through continuous monitoring using the kHealth-Asthma technology which generated insights on causes of asthma symptoms across different seasons. Collectively, they can form the underlying basis for personalized management plan and intervention.

## BACKGROUND AND SIGNIFICANCE
Asthma is a chronic lung inflammatory disease affecting 26 million people in the USA, of which 6 million are children [1]. It is a multifactorial disease with different triggers manifesting as asthma symptoms of various intensities which demands a personalized diagnosis and management plan [2]. Infrequent clinical visits are unable to provide timely feedback and intervention as most of the asthma-exacerbating factors are pollutants from the patient's environment

[3] and lack of medication adherence [4]. Continuous tracking and assessment of a patient's condition, environment, and adherence to a prescribed care plan can improve asthma control and quality of life [5].

While many studies have shown the effectiveness of continuous monitoring, only a few are being evaluated to benefit traditional healthcare practices [6,7]. Propeller Health[8] provides personalized alerts based on inhaler usage and location to primarily improve medication adherence. ENVIROFI [9] and azma.com [10] send a notification to subscribed users when the outdoor environment forecast is poor. Chu et al [11] developed a ubiquitous warning system which sends alerts to healthcare providers based on a patient's location if the outdoor environment is poor. Finkelstein et al [12] developed a web-based approach that captures Forced Vital Capacity test and asthma symptoms from patients and sends alert to hospitals when these parameters are abnormal. AsthmaGuide [13], a home management ecosystem, enables doctors to observe the correlation between symptoms and environmental data. They have classified wheezing sounds as asthmatic wheezing and non-asthmatic wheezing. They also send personalized alerts to patients based on pollen and air quality forecast, but no causal relationships are identified. However, while a number of factors has been shown to influence triggers and control level for an individual asthma patient, monitoring and analyzing diverse data directly relevant to an individual patient has not been adequately evaluated.

Using the kHealth-Asthma technology [14], we monitored and collected diverse data for a cohort of pediatric patients receiving asthma care at the Dayton Children's Hospital (DCH). This paper presents cohort level preliminary data analysis for patients deployed in each of the seasons to identify the major contributors to asthma symptoms. In addition to that, one patient was chosen from each season to illustrate the personalized trigger determination by gathering anecdotal evidence.

**MATERIALS AND METHODS**
kHealth is a framework to personalize digital health by collecting and analyzing multimodal data that complement data collected during routine clinical care, specifically Patient Generated Health data using mobile app and sensors as well environmental data. It is designed to assist self-monitoring and self-appraisal of asthma care with an intent to incorporate self-management, prediction, and intervention of asthma progression [15]. kHealth-Asthma is the adaptation of kHealth for asthma, and comprises of three components: kHealth kit, kHealth cloud, and kHealth Dashboard. The study design including these components, their use for data collection and the data analysis of the 97 patient cohort are discussed next. Other applications for which kHealth has been adapted include post-bariatric surgery monitoring, post-surgery monitoring of Acute Decompensated Heart Failure (ADHF), and dementia.

**kHealth Components**

kHealth Kit
The kHealth kit components are shown in Figure 1. The questionnaire presented by the mobile application on the tablet collects the following data: (i) six types of symptoms: cough, wheeze, chest tightness, hard and fast breathing, can't talk in full sentences, and nose opens wide [16], (ii) medication intake (rescue inhaler and controller medication) with yes or no option, (iii) night-time awakenings due to asthma symptoms, and (iv) activity limitation due to asthma symptoms. The symptoms and medications are collected twice a day, and night-time awakenings and activity limitation are collected once a day (details in Appendix 1). Furthermore, Fitbit is used to collect more granular data for sleep and activity [17]. The lung function measurements (PEF and FEV1) are recorded by the Microlife peak flow meter [18] twice every day. For a given patient's zip code, outdoor environmental parameters are collected at different intervals -- pollen is collected every 12 hours, Particulate Matter (PM2.5), ozone, temperature, and humidity are collected every hour. Pollen is collected from pollen.com [19], PM2.5 and ozone from EPA AIRNow [20], and temperature and humidity from Weather Underground [21]. Foobot collects indoor temperature, humidity, particulate matter, volatile compounds, carbon dioxide, and global pollution index every 5 minutes.

Available literature have evaluated quality and suitability of Fitbit [22,23]. The feasibility study for Foobot has been conducted on our own[24].

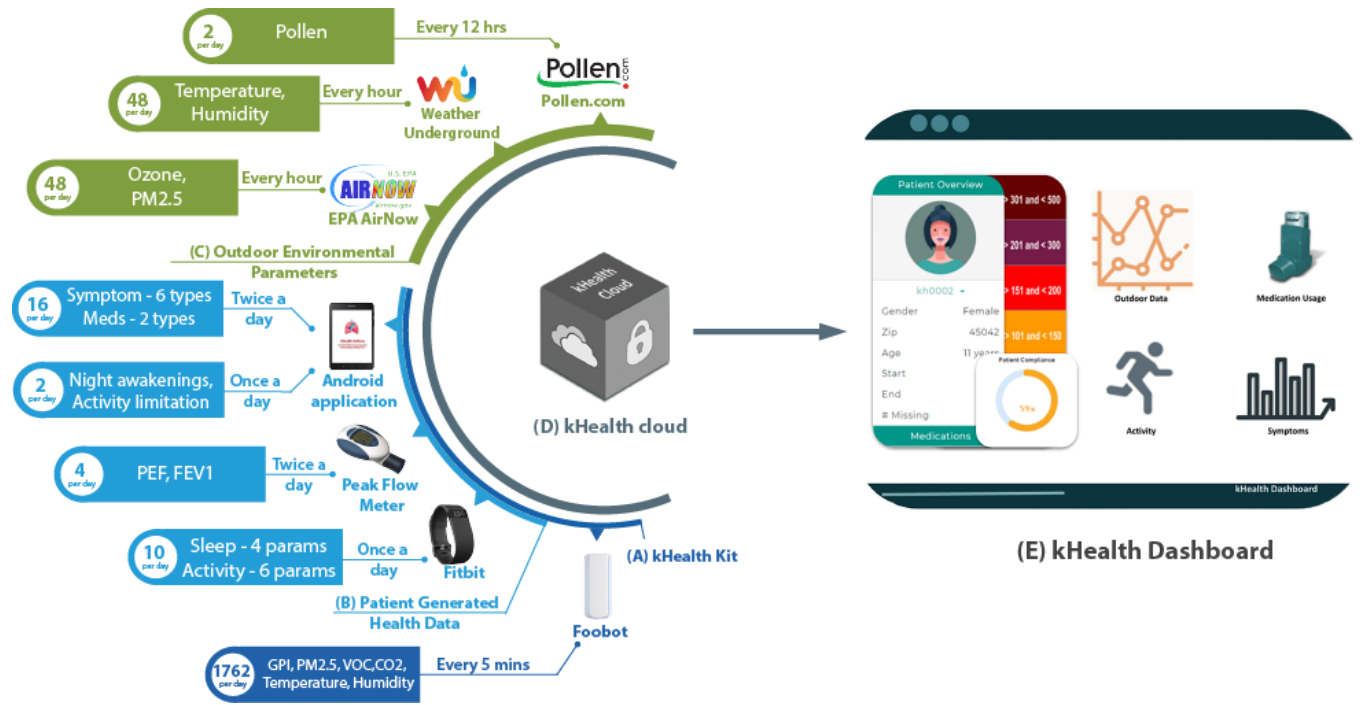

**Figure 1:** The kHealth framework with kHealth kit, kHealth cloud and kHealth Dashboard, showing the frequency of data collection, the number of parameters collected and the total number of data points collected per day per patient. (A-Dark Blue): The kHealth kit components that are given to patients, (B- Light Blue): The kHealth kit components which collects patient generated health data, (C - Green): The outdoor environmental parameters and their sources, (D - Grey): The kHealth cloud, (E): The kHealth Dashboard. Throughout the kHealth ecosystem, all data is anonymized and associated with respective randomly assigned patient IDs.

kHealth Cloud
The multimodal data collected from various sources are brought together on the secure kHealth cloud store. The data about outdoor environment, indoor air quality, activity, and sleep data are collected from their respective API server and stored directly in the kHealth cloud. The data collected using the kHealth application, which includes the patient's symptoms, medication intake, PEF and FEV1 readings, are synced in real-time with Firebase, a Google cloud database [25]. Firebase provides active data listener for client-side that offers data persistence over the network failure and resyncs to the cloud when the network is restored. Data security is maintained in Firebase using a set of data access rules and user authentication. Data synced to Firebase are then fetched and stored into the kHealth cloud. This process forms a pipeline for seamless data streaming from the kHealth application to the kHealth cloud. All of the data stored in the kHealth cloud are then available to Kno.e.sis researchers and clinicians for real-time analysis. Each patient's identity is anonymized by the nurse who consents the patient; no patient-identifiable data is stored anywhere in the kHealth framework outside the clinical setting.

kHealth Dashboard
Since the kHealth kit collects multimodal data at different frequencies, integration and visualization of these data are essential to derive useful insights. kHealth Dashboard [26] (Figure 2), is a visualization and analysis tool designed for use by a clinician or a researcher to review individual and aggregated data, and to explore the potential causal relationship between patient's asthma symptoms and their environments. With real-time data available in the

kHealth cloud, kHealth Dashboard allows real-time monitoring of a patient's asthma condition. This granularity of data presents the clinician with a better picture about patient's asthma condition compared to traditional episodic clinical visits.

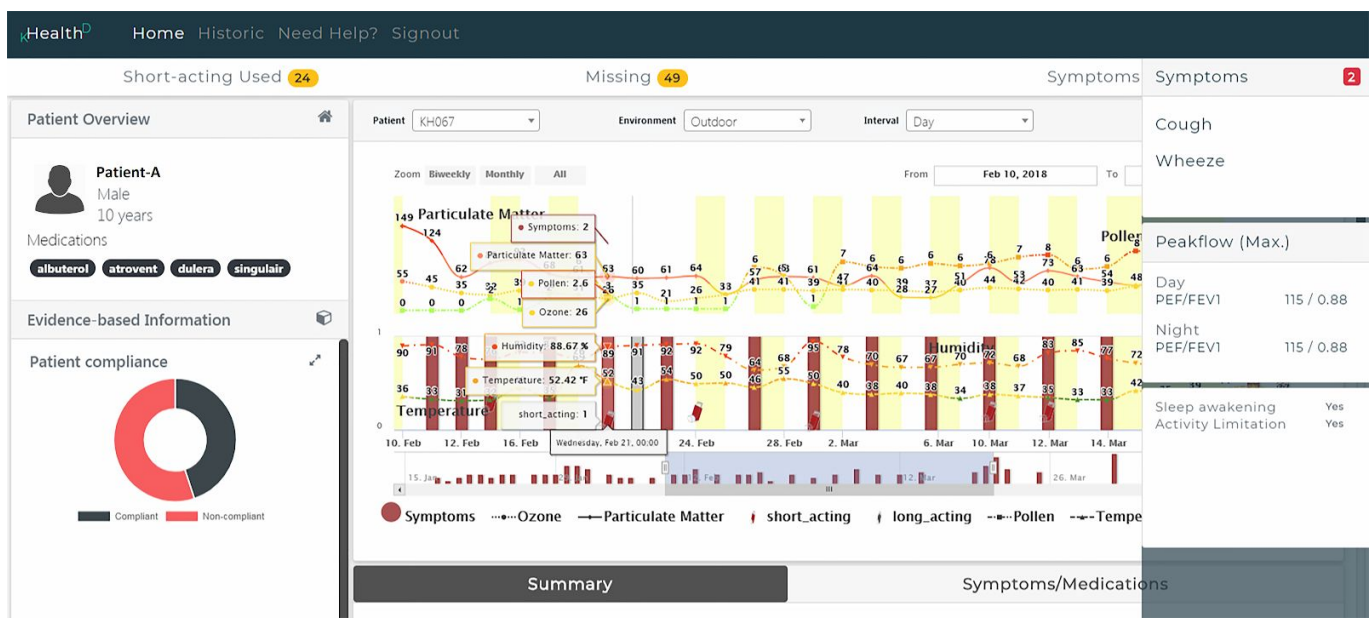

**Figure 2: A screenshot of the kHealth Dashboard visualizing Patient-A's data**

**Study Design and Participants**
Pediatric asthma patients between 5 and 17 years of age receiving care at DCH were consented (with a parent/guardian for very young children) and recruited by a DCH nurse coordinator under the guidance of the clinician. Patient (and parent/guardian) were given training with demonstration and instruction materials. The tablet with Android application, Peak Flow meter, Fitbit and Foobot are given to patients and based on the patient's zip code, outdoor environmental data are collected. The part of data collection, which requires the patient to be actively involved such as Android app questionnaire and Peak Flow meter readings, is referred to as active sensing. The data collection from Fitbit, Foobot and outdoor web services, which do not require active patient involvement, is referred to as passive sensing.

Of the 100 patients consented in the ongoing trial at the time of this manuscript submission, 97 patients had completed the trial and 3 were in progress. Out of 97 patients, 80 were recruited for a month and 17 were recruited for 3 month period. The one month study was designed to validate the efficacy of this method and three month study was included to focus on intervention as three month is the standard duration between two clinical visits to adjust medication based on control level. The enrollment of a patient for one or three months depends on their willingness. This HIPAA compliant study has been approved by DCH's institutional review board. While the deployments started in December 2016 and is still ongoing, the data from December 2016 to July 2018 is included in the analysis. 21 patients were excluded from the analysis due to insufficient active sensing (less 20%), allowing us to analyze data from the remaining patients (n=76).

**RESULT AND DISCUSSION**
Henceforth, the six asthma symptoms (cough, wheeze, chest tightness, hard and fast breathing, can't talk in full sentences, nose opens wide), night-time awakenings, activity limitation, rescue medication intake, and abnormal

PEF/FEV1 value are collectively referred to as *asthma episodes*. Specifically, asthma can cause lung function to decline and can manifest as lower values for PEF and FEV1 parameters [27]. Therefore, the reduction of PEF and FEV1 values beyond 1 standard deviation of the mean were treated as episodes of asthma. The duration of the seasons has been chosen to aid the analysis based on the historical pollen data (Figure 3). The results of cohort level data analysis are shown in Figure 3 for the detection of a wide variety of triggers and causes of asthma episodes.

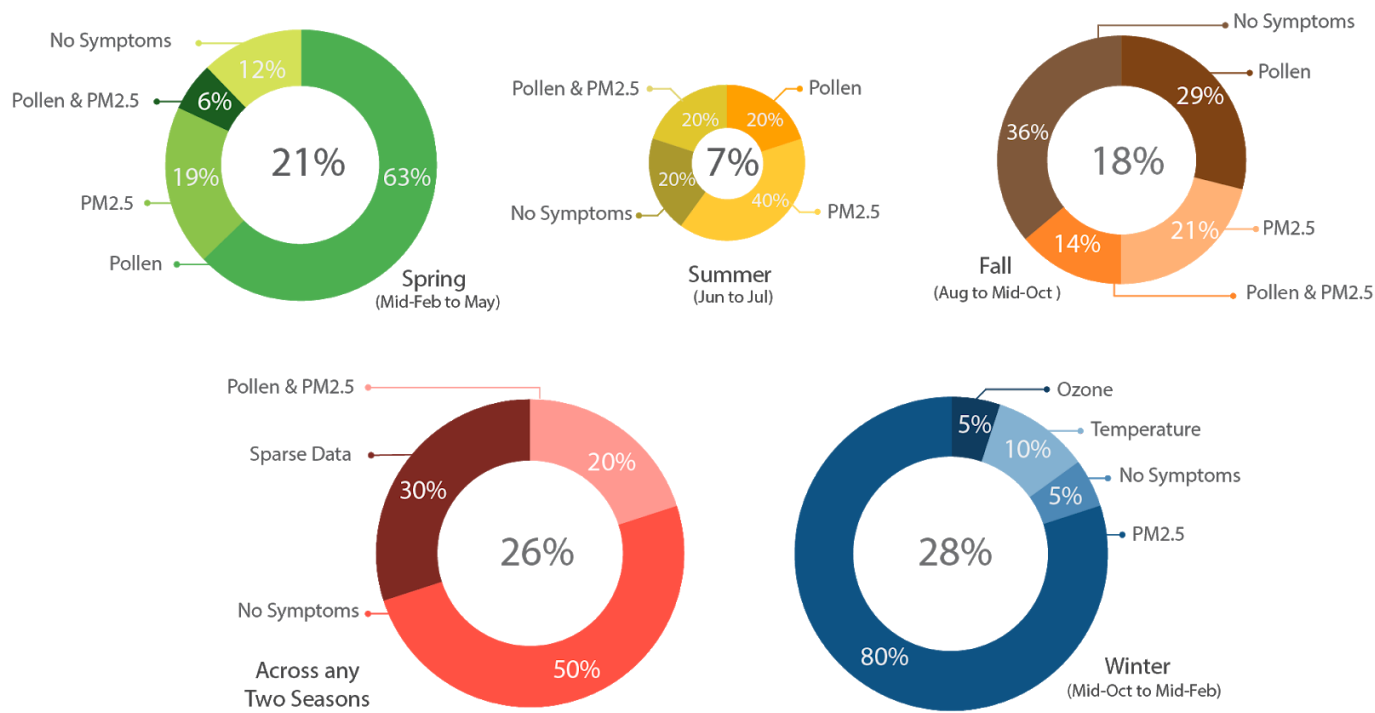

**Figure 3: Personalized (Major) Triggers at the Cohort Level**

One patient with poorly controlled asthma was chosen for each season for illustration. As such, two cases containing adequate data but poor kit compliance have been included. Patient-A's deployment period straddled two seasons, winter to spring, which permitted the study of asthma's behavior both in the presence and in the absence of pollen. The other three patients were exclusively within one season: Patient-B in winter, Patient-C in fall, and Patient-D in summer. The days the patients did not answer the questionnaire were excluded from the analysis. Each patient's deployment period was divided into two: a learning period and a prediction period. The data from the learning period was used to learn the causal association between the patient's asthma episodes and triggers. The data from the prediction period was used to evaluate the ability to predict the likelihood of an episode based on the presence of the triggers. Ultimately, once a reliable personalized model associating triggers with asthma episode is attained, it can be used to guide plausible action plan including preventive or remedial measures, as well as further targeted evaluations to help develop better personalized care.

Maximum values of outdoor environmental data (ozone, PM2.5) over a day were considered to identify the correlation between triggers and asthma episodes. The healthy range for each outdoor parameters are -- Pollen [28]: 0 to 2.4, Ozone and PM2.5 [29]: 0 to 50. Any value above or below the healthy range on the day with asthma episodes, or the previous day, is counted as a contributor to the patient's asthma episodes. While we collected extensive Foobot data, the results on the influence of indoor environment are inconclusive, potentially because of a number of deficiencies in our instructions such as (a) clarity on the placement of Foobot, and (b) potential for electrical interference, when some patients did not power the device as required by the manufacturer. In the study results recorded thus far, the outside environment provides more reliable signals for asthma control. Further, the sleep and activity data from Fitbit did not provide corroborative evidence for asthma signs due to several confounding factors. In contrast, self-reported data obtained through the Android app based questionnaire proved to be more reliable. We analyze concrete patient cases to obtain insights about asthma triggers, patient behavior and their condition from evidences collected by kHealth-Asthma technology in different seasons.

**Winter to Spring**
Patient-A was diagnosed with severe asthma, and monitored for 13 weeks encompassing winter to spring 2018 and answered the questionnaire for 46 days. The patient was prescribed Albuterol and Atrovent (as rescue medication), and Dulera and Singulair (as controller medication). The patient took rescue medication for 24 days, experienced cough, wheeze or chest tightness for 39 days, had limited activity for 26 days, abnormal PEF/FEV1 for 2 days and was 15% compliant in taking controller medication.

As pollen was absent in the first half of the deployment period and present in the second half, the deployment period was divided and evaluated for two periods: first 6 weeks without pollen, and last 7 weeks with pollen. The patient's reaction to ozone and PM2.5 manifesting as asthma episodes were also studied separately during these two periods to disambiguate among multiple triggers. The 6 weeks without pollen was further divided into two sub-periods: first 4 weeks as the learning period and rest (2 weeks) as the prediction period. Similarly, the 7 weeks without pollen was further divided into 4 weeks as the learning period and 3 weeks as the prediction period.

Table 1. Learning causal associations and Prediction of Asthma episodes for Patient-A

| **Absence of Pollen** First 6 weeks | | | | **Presence of Pollen** Rest of the 7 weeks | | | |
|---|---|---|---|---|---|---|---|
| **Learning** 4 weeks | | **Prediction** 2 weeks | | **Learning** 4 weeks | | **Prediction** 3 weeks | |
| **Pollen** | 0 days | **Pollen** | 0 days | **Pollen** | 17 days | **Pollen** | 3 days |
| **PM2.5** | 20 days | **PM2.5** | 5 days | **PM2.5** | 14 days | **PM2.5** | 2 days |
| **Ozone** | 1 day | **Ozone** | 0 days | **Ozone** | 0 days | **Ozone** | 1 day |
| **Asthma episodes** | 21 days | **Asthma episodes** | 5 days | **Asthma episodes** | 17 days | **Asthma episodes** | 3 days |
| **Outdoor Temperature:** 7 to 55 F | | | | **Outdoor Temperature:** 30 to 75F | | | |
| **Outdoor Humidity:** 20 to 98 % | | | | **Outdoor Humidity:** 20 to 99% | | | |

In the absence of pollen, PM2.5 appeared as the major contributor of asthma episodes (shown in Table 1). During the prediction period, PM2.5 was high on the days the patient reported symptoms. It can be inferred that PM2.5 is one of the contributors of the patient's asthma episodes as other triggers were in the healthy range. Similarly, in the presence of pollen, pollen was the major contributor and PM2.5 was the second major contributor. During the prediction period, both pollen and PM2.5 were high on almost all the days the patient reported asthma episodes. Though the patient showed asthma episodes both in the presence and in the absence of pollen, the patient started experiencing severe symptoms only in the presence of pollen. The patient experienced 6 out of 9 occurrences of chest tightness (chest tightness is more severe than cough or wheeze) and 6 out of 7 occurrences of night awakenings in the presence of pollen. For this patient, the presence of pollen or PM2.5 in the unhealthy range appears to be prime contributing factor for asthma episodes, and the combined presence is associated with increased intensity of asthma episodes. When validated with the clinician, the patient was identified allergic to pollen using skin test. The Asthma Control Test scores before and after the deployment confirm that the patient's asthma control was suboptimal.

Through continuous monitoring, we found that PM2.5 and pollen were the contributors to the patient's asthma episodes, and the patient had poor adherence to controller medication. To improve asthma management, the intervention can be personalized by alerting patient about high pollen and PM2.5 forecast. Further, notification can be sent to improve the compliance for controller medication. If asthma episodes recur even after being compliant toward controller medication, the clinician can intervene with the modified asthma action plan.

**Winter**

Patient-B was classified as having moderate asthma and monitored for 13 weeks in winter 2017 to 2018. But data from only the first 9 weeks was available for analysis as the patient didn't answer for 4 weeks towards the end of the deployment. The patient was prescribed Albuterol (as rescue medication) and Singulair (as controller medication), and answered the questionnaire for 50 days of which the patient experienced asthma episodes for 45 days. Patient-B experienced wheezing for 27 days, activity limitation for 15 days, night-time awakenings on 1 day, took rescue medication for 24 days, had abnormal PEF/FEV1 values for 6 days, and was 50% compliant toward controller medication. Since the environment was uniform throughout the deployment, first 6 weeks were used to determine the triggers, and the last 3 weeks were used for prediction.

Table 2. Learning causal associations and Prediction of Asthma episodes for Patient-B

| **Learning** 6 weeks | | **Prediction** 3 weeks | |
|---|---|---|---|
| **Pollen** | 0 days | **Pollen** | 0 days |
| **PM2.5** | 21 days | **PM2.5** | 19 days |
| **Ozone** | 0 days | **Ozone** | 0 days |
| **Asthma episodes** | 24 days | **Asthma episodes** | 21 days |
| **Outdoor Temperature:** 19 F to 60 F | | **Outdoor Temperature:** -2 F to 58 F | |
| **Outdoor Humidity:** 17% to 99% | | **Outdoor Humidity:** 25% to 99% | |

As shown in Table 2, PM2.5 was found to be the major contributor. When the findings were applied to predict the asthma episodes, PM2.5 was bad for 19 days and patient experienced asthma episodes for 21 days. Specifically, for the 2 days the patient experienced asthma episodes, none of the outdoor parameters were high on that day or the previous day. However, it was observed that PM2.5 was consistently high for a week before the 2-3 days of asthma episodes. The effect of prolonged exposure to high PM 2.5 can explain persistent airway hyper-reactivity and thereby asthma episodes. It is desirable to continue further evaluations to understand the triggers in other seasons.

In consequence, the patient can be proactively notified when PM2.5 is forecasted to be in the unhealthy range and may also be reminded to take prescribed medication. To enhance the asthma management, the patient can improve compliance toward controller medication and avoid exposure to PM2.5 when it is high. If this does not help, the patient should be reevaluated to adjust the asthma control plan.

**Fall**

Patient-C was diagnosed with moderate asthma and monitored for 5 weeks and 4 days in the fall of 2017. The patient answered the questionnaire 33 out of 36 days of deployment, of which the patient showed asthma episodes for 17 days. The patient was prescribed Albuterol (as rescue medication), Symbicort and Singulair (as controller medications), and Prednisone (oral steroid). The patient had cough and wheeze for 11 days, had activity limitation for 4 days, took rescue medication for 6 days, showed abnormal PEF/FEV1 for 9 days, and had compliance toward controller medication of 63%. The outdoor environment remained uniform with respect to pollen, ozone, and PM2.5 throughout the deployment. First 4 weeks of data were used for learning and the rest were used for prediction.

Table 3. Learning causal associations and Prediction of Asthma episodes for Patient-C

| Learning Period 4 weeks | | Prediction Period 1 week and 4 days | |
|---|---|---|---|
| **Pollen** | 11 days | **Pollen** | 10 days |
| **PM2.5** | 8 days | **PM2.5** | 10 days |
| **Ozone** | 2 days | **Ozone** | 6 days |
| **Asthma episodes** | 12 days | **Asthma episodes** | 5 days |
| **Outdoor Temperature:** 65 F to 85 F | | **Outdoor Temperature:** 55 F to 80 F | |
| **Outdoor Humidity:** 70% to 90% | | **Outdoor Humidity:** 50% to 98% | |

For this patient, pollen was found to be the major contributor and PM2.5 was the second major contributor (shown in Table 3). But the patient experienced asthma episodes only for 5 days in the prediction period. This could be both due to the improved compliance with controller medication up from 44% in the learning period to 100% in the prediction period, and improved control of airway inflammation by oral steroid intake for a week in the learning period.

kHealth system has the potential to aid Patient-C in self-management of asthma by alerting the patient when pollen or PM2.5 can exacerbate asthma and reminding the patient to take medication to improve compliance. Further, the clinician can also be notified when the patient takes the oral steroid, an indication of poor asthma control, to enable timely intervention. To exonerate some triggers and identify triggers precisely, the experiment should be repeated in the season when pollen is absent.

**Summer**

Patient-D has mild asthma, and was monitored for 4 weeks and 3 days in the summer of 2018. The patient was prescribed Albuterol (as rescue medication), Asmanex and Singulair (as controller medications). Patient-D has answered the questionnaire for 29 out of 30 days and experienced asthma episodes on 6 days. The patient experienced cough, wheeze, chest tightness, or hard and fast breathing for 5 days, activity limitation on 1 day, night-time awakening on 1 day, and the compliance toward the controller medication was 70%. The lung function measurements were normal throughout the study period. Out of the 6 days the patient showed asthma episodes, pollen was in an unhealthy range for 3 days, ozone for 4 days, and PM2.5 on 6 days. Based on the observation, PM2.5 is suspected to be the major contributor, followed by ozone and pollen. However, this patient had sparse asthma episodes to build a predictive model for the asthma behavior. In general, none of the patients deployed in summer had sufficient asthma episodes to learn a predictive model for the positive triggers of the episodes.

**CONCLUSION**

The infrequent clinical visit as practiced by traditional healthcare protocol is unable to provide timely feedback and enable intervention. Through continuous monitoring, kHealth kit is able to provide detailed insights to the clinician about the personalized triggers for asthma patients and their compliance toward the prescribed asthma control protocol. Specifically, for patients like patient-A that spanned two seasons, the kHealth kit determined the triggers precisely along with the patient's adherence to the prescribed asthma action plan. This can aid the clinician in tailoring the asthma control protocols for a patient leading to better asthma management. Further kHealth kit was able to capture triggers across different seasons, which was evident from the determination of the variety of personalized triggers for 4 example patients chosen across the different seasons.

**FUTURE WORK**

We plan to repeat the observational trial across different seasons for each patient. Of 76 patients, 27% (n=20) of the patients monitored in winter, and 29% (n=22) of the patients monitored in fall and in spring experienced asthma episodes. Redeployment will be carried out for winter patients to discover potential triggers in other seasons, and also for spring and fall patients to disambiguate among multiple triggers by exonerating some. For 9% (n=7) of the patients, the deployment spanned across two seasons, yielding sparse data and hence insufficient for any reliable conclusion. Deployment for this subgroup will be attempted again, and one month deployments that span across two seasons will be avoided in future. 25% (n=19) of the entire patient cohort didn't experience any asthma episodes. Eventually, with the kHealth system, we expect to identify triggers across seasons for each patient that cause worsening of the patient's asthma and aid the clinician with insights about triggers and patient compliance for personalized action plan. This study has also provided insights that can help design future study involving self management and intervention. For such a study, some of the less important data types will be dropped and a version of the kHealth Dashboard will be developed for use by patients that has customizable functionality, look and feel.

**COMPETING INTERESTS**

The authors declare no competing financial interests.

**CONTRIBUTORS**

AS is the PI of the kHealth-Asthma project, MK and KT are investigators. AS conceptualized and initiated kHealth technology and its applications. AS and KT led the study design, MK lead all clinical aspects. The significance of the student contributions is in the order RV, DK, HY and UJ. RV analyzed the data and lead the manuscript writing. DK and HY aided the data analysis. DK developed the cloud services and wrote kHealth cloud and kHealth Dashboard section of the manuscript, DK and RV were involved in development of kHealth Dash, HY coordinated the data collection and DK, RV and UJ were involved, UJ supervised the implementation of Android application and validated the sensors. RV wrote the initial manuscript with guidance of KT, all authors reviewed and edited the manuscript.

**ACKNOWLEDGEMENTS**

We thank Dr. Tanvi Banerjee and Vaikunth Sridharan for their contributions. Dr. Tanvi Banerjee has been involved in the early part of the kHealth-Asthma project as an investigator. Vaikunth Sridharan was the primary implementer of the kHealth Dashboard and completed his MS Thesis that focused on its development.

**FUNDING**
This research is supported by NICHD/NIH under the Grant Number: 1R01HD087132. The content of this paper is solely the responsibility of the authors and does not necessarily represent the official views of the NIH.

**SUPPLEMENTARY MATERIAL**
Supplementary material is available at Journal of the American Medical Informatics Association online.

**LEGEND**

Figure 1: The kHealth framework with kHealth kit, kHealth cloud and kHealth Dashboard, showing the frequency of data collection, the number of parameters collected and the total number of data points collected per day per patient. (A-Dark Blue): The kHealth kit components that are given to patients, (B- Light Blue): The kHealth kit components which collects patient generated health data, (C - Green): The outdoor environmental parameters and their sources, (D - Grey): The kHealth cloud, (E): The kHealth Dashboard. Throughout the kHealth ecosystem, all data is anonymized and associated with respective randomly assigned patient IDs.

Figure 2: A screenshot of the kHealth Dashboard visualizing Patient-A's data

Figure 3: Personalized (Major) Triggers at the Cohort Level

Table 1. Learning causal associations and Prediction of Asthma episodes for Patient-A

Table 2. Learning causal associations and Prediction of Asthma episodes for Patient-B

Table 3. Learning causal associations and Prediction of Asthma episodes for Patient-C

______________________________________________________________________

**SUPPLEMENTARY TEXT**
**Appendix 1: The questionnaire from kHealth Asthma Android application**

| Question | Answer type |
| --- | --- |
| Did you experience any of the asthma symptoms? | Cough, Wheeze, Chest tightness, Hard and Fast breathing, Can't talk in full sentences, Nose opens wide |
| How many times did you take *[name of the rescue inhaler prescribed]* inhaler today due to asthma symptoms? | 1,2,3,4,5,6+ |
| Did you take *[name of the controller medication prescribed]* today? | Yes, No |
| How much did asthma symptoms limit your activity? | None, A little, Most of the day, At least half of the day |
| Did you wake up last night due to asthma symptoms? | Yes/No |